\documentclass{article}
\usepackage{spconf}

\usepackage{amsmath, bbm}
\usepackage{amssymb}
\usepackage{amsfonts}
\usepackage{amsthm}
\usepackage{algorithm}
\usepackage{algorithmic}
\usepackage{nicefrac}
\usepackage{bm}
\usepackage{hyperref}
\usepackage{cite}
\usepackage[utf8]{inputenc} 
\usepackage[T1]{fontenc}
\usepackage{graphicx}
\usepackage[font=small]{caption}
\usepackage{tikz}
\usetikzlibrary{positioning}
\usepackage{pgfplots}
\usepackage{epstopdf}
\usepackage[labelformat=simple]{subcaption}
\usepackage{multirow}
\usepackage{lipsum}
\usepackage{mathtools}

\newtheorem{Problem}{Problem}

\newcommand{\update}[1]{#1}

\input{mysymbol.sty}

\title{Delay-Oriented Distributed Scheduling Using Graph Neural Networks}

\name{Zhongyuan Zhao$^\star$, Gunjan Verma$^\dag$, Ananthram Swami$^\dag$, and Santiago Segarra$^\star$
\thanks{Research was sponsored by the Army Research Office and was accomplished under Cooperative Agreement Number W911NF-19-2-0269. 
		The views and conclusions contained in this document are those of the authors and should not be interpreted as representing the official policies, either expressed or implied, of the Army Research Office or the U.S. Government. 
		The U.S. Government is authorized to reproduce and distribute reprints for Government purposes notwithstanding any copyright notation herein.
		\newline
		Emails:  \{zhongyuan.zhao, segarra\}@rice.edu, \{gunjan.verma.civ,  ananthram.swami.civ\}@army.mil. }}
\address{$^\star$Rice University, USA  \hspace{1cm} $^\dag$US Army’s DEVCOM Army Research Laboratory, USA}

\begin{document}
\ninept
\renewcommand{\baselinestretch}{0.95}
\maketitle
\begin{abstract}
In wireless multi-hop networks, delay is an important metric for many applications. 
However, the max-weight scheduling algorithms in the literature typically focus on instantaneous optimality, in which the schedule is selected by solving a maximum weighted independent set (MWIS) problem on the interference graph at each time slot. 
These myopic policies perform poorly in delay-oriented scheduling, in which the dependency between the current backlogs of the network and the schedule of the previous time slot needs to be considered. 
To address this issue, we propose a delay-oriented distributed scheduler based on graph convolutional networks (GCNs). 
In a nutshell, a trainable GCN module generates node embeddings that capture the network topology as well as multi-step lookahead backlogs, before calling a distributed greedy MWIS solver.
In small- to medium-sized wireless networks with \update{heterogeneous transmit power, where a few central links have many interfering neighbors}, our proposed distributed scheduler can outperform the myopic schedulers based on greedy and instantaneously optimal MWIS solvers, with good generalizability across graph models and minimal increase in communication complexity.  
\end{abstract}
\begin{keywords}
Maximum weighted independent set, graph neural networks, distributed scheduling, latency.
\end{keywords}
\section{Introduction}\label{sec:intro}

Wireless multi-hop networks are fundamental to modern wireless communications, including military communications, wireless backhaul for 5G and beyond, and Internet of Things (IoT) \cite{Lin06,sarkar2013ad,kott2016internet,akyildiz20206g}.
One challenge of wireless multi-hop networks is distributed resource allocation, such as link scheduling, without the help of infrastructure.
Specifically, link scheduling determines which links should transmit and when should they transmit, along with other relevant parameters~\cite{Joo09,marques2011optimal}. 
In this paper, we focus on link scheduling in wireless networks with time-slotted orthogonal multiple access, in which a time slot comprises a scheduling phase followed by a transmission phase \cite{Kabbani07,paschalidis2015message}. 
The optimal scheduling problem in wireless multi-hop networks is typically formulated as solving a maximum weighted independent set (MWIS) problem on a \emph{conflict graph}~\cite{basagni2001finding,Kabbani07,Joo09,joo2012local,joo2015distributed,marques2011optimal,sanghavi2009message,du2016new,Li18Ising,Douik2018,paschalidis2015message,joo2010complexity,cheng2009complexity,zhao2021icassp,zhao2021jstsp},
in which a vertex represents a link in the wireless network, an edge captures the interference relationship between two links, and the vertex weight is the utility of the corresponding link.
The scheduling scheme contains two main parts: 1)~a per-link utility function to evaluate the importance of a link relative to the scheduling objective, and 2)~an approximate (and possibly distributed) solver for the associated MWIS problem, which is known to be NP-hard~\cite{cheng2009complexity,joo2010complexity}.

\begin{figure}
    \centering
    \vspace{-3mm}
    \begin{subfigure}[b]{0.45\linewidth}
    \includegraphics[height=0.9in]{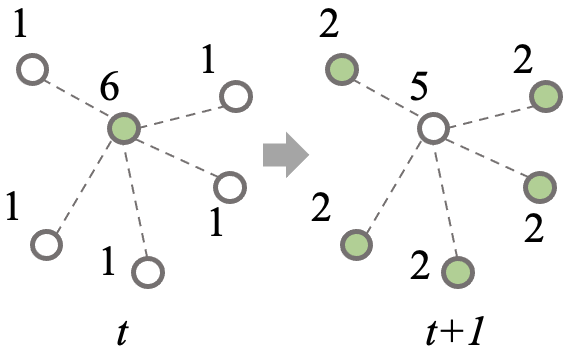}
    		\caption{}
    \label{fig:motivation:opt}
	\end{subfigure}%
	   \begin{subfigure}[b]{0.45\linewidth}
    \includegraphics[height=0.9in]{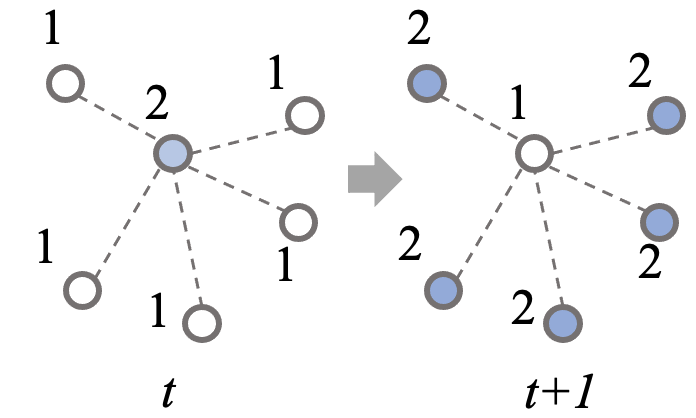}
    		\caption{}
    \label{fig:motivation:greedy}
	\end{subfigure}%
    \vspace{-4mm}
    \caption{The steady-state transitions in a toy conflict graph with star topology, constant arrival rate of 1, link rate of 2, utility function given by the queue length, and initial state of all empty queues. 
    The colored nodes are scheduled by
    (a) an optimal MWIS scheduler and (b) a greedy scheduler. 
    The latter has shorter average delay.}
    \vspace{-6mm}
    \label{fig:motivation}
\end{figure}

Although the MWIS-formulated schedulers seek to maximize
the total throughput or utility \cite{basagni2001finding,joo2012local,joo2015distributed,marques2011optimal,Kabbani07,Joo09,paschalidis2015message}, they usually have poor delay performance~\cite{xue2012delay}.
Indeed, the memoryless nature of the \update{queue-based utility functions \cite{basagni2001finding,Joo09,joo2012local,joo2015distributed,marques2011optimal,Kabbani07,paschalidis2015message}} makes it difficult to optimize network metrics that depend on serial decisions such as delay. 
As an example, consider a conflict graph with star topology for a wireless network with 6 links (represented as the vertices of the graph) as in Fig.~\ref{fig:motivation}.
Furthermore, consider \update{an initial state of all empty queues,} a constant arrival rate of $1$ and a link rate of $2$ (both in packets per time slot) for each link, and set the per-link utility as the queue length so that links with longer queues are preferentially scheduled. 
Since the star has two maximal independent sets (one is the central node, the second is all peripheral nodes), the optimal MWIS scheduler will schedule the set that achieves the maximum sum of queue lengths, leading to the \update{alternation} of two network states as in Fig.~\ref{fig:motivation:opt}. 
The greedy scheduler builds the schedule iteratively by adding one node at a time starting with the one with the largest queue and subsequently remove its neighbors to avoid collisions, leading to the alternation of the two network states in Fig.~\ref{fig:motivation:greedy}.
The example reveals that the average queue length under the optimal MWIS scheduling is $\overline{q}_{opt}=2.17$, which is greater than that of the greedy scheduler, $\overline{q}_{Gr}=1.5$. 
Consequently, for unsaturated network traffic, the optimal MWIS solver has poorer delay performance than a simple greedy heuristic.

Existing approaches to reduce the delay of scheduling 
include setting delay as a constraint of optimization \cite{Jaramillo2011scheduling,hou2010utility}, replacing the queue length in the conventional utility functions \cite{basagni2001finding,Joo09,joo2012local,joo2015distributed,marques2011optimal,Kabbani07,paschalidis2015message} with metrics of delay (e.g., virtual queues of congestion  \cite{xue2012delay}, sojourn time \cite{hai2018delay}, age-of-information \cite{Hsu2017age}), and machine learning-based solutions \cite{gupta2020learning,lee2021graph,gao2017multi,wang2021learning}. 
Conventional approaches \cite{Jaramillo2011scheduling,hou2010utility,xue2012delay,hai2018delay,Hsu2017age} generally do not fully consider network topology. 
In contrast, machine learning-based approaches incorporate topological information as part of the network state either implicitly through, e.g., a multi-layer perceptron (MLP) tied to a specific topology~\cite{gao2017multi}, or explicitly through graph embedding \cite{lee2021graph} and graph neural networks \cite{wang2021learning}.
Schedules are generated either by algorithmic frameworks based on the outputs of neural networks, such as utility \cite{gupta2020learning} and bias \cite{gao2017multi}, or directly by MLP-based binary classifiers~\cite{lee2021graph}.
However, these machine learning-based approaches are either limited to centralized scheduling~\cite{gupta2020learning,lee2021graph}, continuous-valued problems \cite{wang2021learning} (link scheduling is discrete-valued), or have poor scalability \cite{gao2017multi}.

In this paper, we depart from existing approaches and propose a scheme based on graph convolutional networks (GCNs) \cite{kipf2016semi} and inspired by deep Q learning (DQL), denoted as GCN-DQL, which can improve the delay performance of wireless networks by generating per-link utilities that are aware of the network state and topology.
Our scheduler is composed of a GCN followed by the application of a {non-differentiable} distributed local greedy solver (LGS)~\cite{joo2012local}, similar to the architecture adopted in~\cite{zhao2021icassp,zhao2021jstsp}.
However, we propose a different training approach to capture future returns of the trained scheduling policy. 
More specifically, we introduce a reward signal to evaluate the relative performance of the trained policy compared to that of the vanilla LGS in a $K$-step lookahead scheduling based on the current state (which includes the backlogs and capacities of links) and topology of the wireless network. 
Although our method relies on centralized training, it can be deployed in a fully distributed manner thanks to the distributed nature of the GCN and~LGS.

\vspace{1mm}
\noindent
{\bf Contribution.} The contributions of this paper are twofold:
1)~we propose the first GCN-based distributed scheduler with temporal lookahead capabilities, and
2)~through numerical experiments, we demonstrate the superior performance of the proposed method as well as its generalizability over different topologies.

\section{System Model and Problem Statement}
\label{sec:problem}

Consider a wireless multi-hop network, where an (undirected) link $(i,j)$ implies that user $i$ and user $j$ can communicate with each other. 
A flow describes the stream of packets from a source node to a destination node. 
A flow may pass through multiple links determined by a routing scheme. 
In each link, there is a queuing system $q$ for packets of all the flows as well as exogenous arrivals.
\update{We denote by $q(v)$ the sum of bidirectional queues for a link $v$.} 

The scheduling algorithm works on the \emph{conflict graph}, $\mathcal{G}(\ccalV,\ccalE)$, which is defined as follows: a vertex $v\in\ccalV$ represents a link in the wireless network, and the presence of an undirected edge $e=(v_a,v_b)\in\ccalE$ captures the interference relationship between links $v_a, v_b \in\ccalV$.
The interference relationship in the system is considered to follow a physical distance model~\cite{cheng2009complexity}. 
Two links interfere with each other if their incident users are within a certain distance such that their simultaneous transmission will cause the outage probability to exceed a prescribed level, or they share the same user with only one radio interface.
For the rest of this paper, we focus on the conflict graph $\mathcal{G}$, which we assume to be known; see, e.g.,~\cite{yang2016learning} for its estimation. 
In principle, the interference zone of each link (hence $\mathcal{G}$) depends on the transmit power and antenna directivity of the corresponding users.
To simplify the analysis and avoid this dependence, we consider the scenario in which all the users transmit at power levels that do not vary with time.

An independent (vertex) set in a graph is a set of nodes such that no two nodes in the set are neighbors of each other.
From the definition of $\mathcal{G}$, only wireless links that form independent sets in $\mathcal{G}$ can communicate simultaneously in time and frequency under orthogonal access.
Our link scheduling aims to minimize the average communication delays over a long time horizon.
More precisely, we describe the \emph{network state} at time $t$ by the tuple $(\mathcal{G}(t), \mathbf{q}(t), \mathbf{r}(t))$ consisting of the conflict graph $\mathcal{G}(t)$ (potentially changing over time), queue lengths $\mathbf{q}(t)$, and link rates $\mathbf{r}(t)$.
If we denote by $\mathcal{C}$ the space of all functions that go from network states into independent sets of the graph, we can formally define our problem as follows. 

\begin{Problem}\label{P:main} 
    For a time horizon of interest $T$, we want to solve for the delay-optimal scheduler given by
    \begin{subequations}\label{eq:utility}
    \vspace{-1mm}
	\begin{align}
	& c^* = \argmin_{c \in \mathcal{C}} \,\, \mathbb{E} \left(\update{ \frac{1}{T+1}\sum_{t=0}^{T} \frac{\lVert\bbq(t)\rVert_{1}}{|\ccalV(t)|} } \right) \label{eq:utility:obj}\\[7pt]
	\text{s.t. } \,\,\,  & \hat{\boldsymbol{v}}(t) = c(\mathcal{G}(t), \mathbf{q}(t), \mathbf{r}(t)), \label{eq:utility:mwis} \\
	& \hspace{-9mm} q_v(t+1) \!=\! \!
	\begin{cases}
	\! q_v(t) \! + \! a_v(t) \quad \qquad \qquad \qquad \quad \,\,\,\,   \text{if} \,\, v \not\in \hat{\boldsymbol{v}}(t), \\
	\! q_v(t) \! + \!  a_v(t) \! - \! \min(r_v(t),q_v(t)) \,\,\, \text{if} \,\, v \in \hat{\boldsymbol{v}}(t), \label{eq:queue_update} 
	\end{cases}
	\end{align}
	where 
	both constraints hold for every time $t=0, \ldots, T$ and the second constraint holds for all $v \in \mathcal{V}$. 
\end{subequations}
\end{Problem}

To better understand Problem~\ref{P:main}, first notice that in constraint~\eqref{eq:utility:mwis} we are defining the set of vertices $\hat{\boldsymbol{v}}(t)$ to be scheduled at every time $t$. 
Since $c \in \mathcal{C}$, these sets of vertices are guaranteed to be independent sets, thus, feasible scheduling choices.
Constraint~\eqref{eq:queue_update} updates the queues at every vertex accordingly. To be precise, if a node $v$ has not been scheduled, its queue at the next time point is given by the previous queue plus any arrivals $a_v(t)$. 
On the other hand, if a node was scheduled then we need to subtract the packets that were sent, which equals the minimum between the queue length at that vertex and the rate achievable.
Among all possible scheduling functions in $\mathcal{C}$ we seek to find the one that minimizes the objective in~\eqref{eq:utility:obj}, which computes the average (over time and over vertices) queue length.
Notice that the queue lengths ultimately depend on the (unknown) link rates and arrival rates.
Thus, we model queues as random variables and we seek to minimize their expected value with respect to the random distributions of arrivals and link rates.
It should be noted that Problem~\ref{P:main} is exclusively focused on the optimal scheduler (implemented at the link layer) and, thus, we have no decision over the arrival rates (possible defined by a routing algorithm implemented at the network layer).

Finding an exact solution to Problem~\ref{P:main} is extremely challenging.
Notice that even in the single-step case ($T=0$) and in the absence of any randomness, selecting the optimal (weighted) independent set is known to be NP-hard~\cite{cheng2009complexity,joo2010complexity}.
Moreover, the optimization in~\eqref{eq:utility} is more challenging than a vanilla MWIS problem for at least three reasons:
i)~We are optimizing over the set of functions $\mathcal{C}$ that goes from network states to independent sets, 
ii)~Our objective depends on random arrivals and link rates, and
iii)~Our objective depends on multiple scheduling instances.
In the next section we present our solution to Problem~\ref{P:main}, which addresses the aforementioned challenges.


\begin{figure*}
	\vspace{-0.1in}
	\centering
		\includegraphics[width=0.8\linewidth]{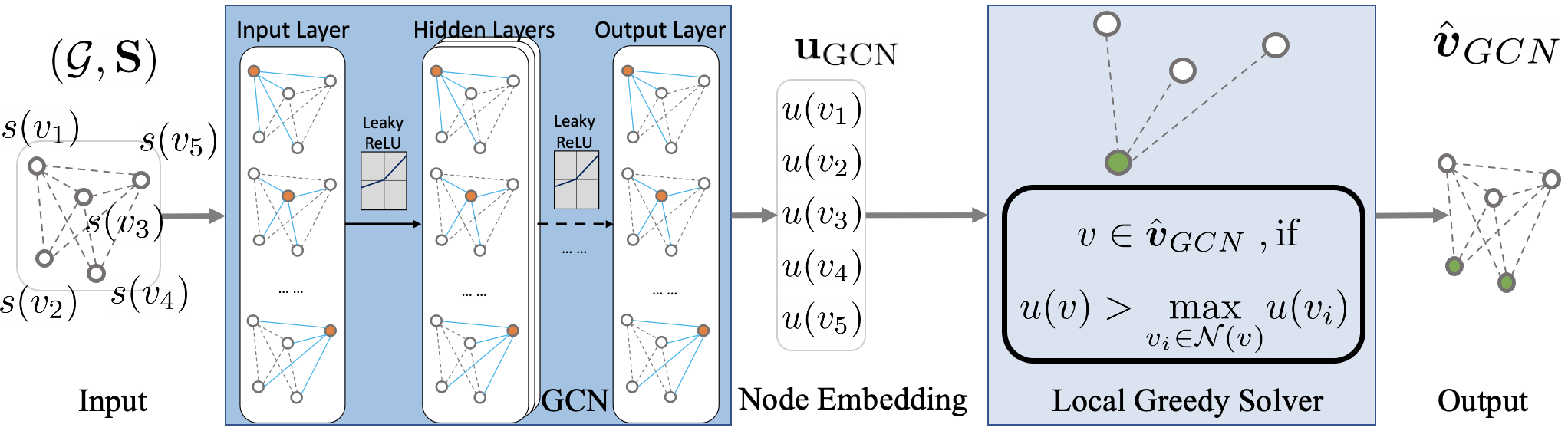}\vspace{-0.05in}
	\caption{Architecture of GCN-based distributed scheduler \cite{zhao2021icassp,zhao2021jstsp}. First, the conflict graph $\ccalG$ and node features $\bbS$ are encoded into the scalar embeddings $\bbu_{\mathrm{GCN}}$ via a GCN. Then, the node embeddings $\bbu_{\mathrm{GCN}}$ are fed into a distributed local greedy solver to generate the solution $\hat{\boldsymbol{v}}_{\mathrm{GCN}}$.} \label{fig:system}
	\vspace{-0.1in}
\end{figure*}

\section{Delay-oriented scheduling with GCN}
\label{sec:solution}

To address the three challenges in solving Problem~\ref{P:main}, we respectively employ three strategies:
i)~Defining a parameterization of a subspace of $\mathcal{C}$ and optimizing over those parameters,
ii)~Training from \update{collected} historic data whose arrivals and link rates follow the distributions of interest, and 
iii)~Incorporating a lookahead reward in our training that penalizes greedy scheduling choices.

To ensure that the output of our scheduler is a valid independent set, our approach consists of two steps: a trainable GCN~\cite{kipf2016semi} that learns per-link utilities $\bbu_{\mathrm{GCN}}$ followed by the application of classical LGS~\cite{joo2012local}.
More precisely, if we omit $t$ for notational simplicity and denote by $\hat{\boldsymbol{v}} = \mathrm{LGS}(\ccalG, \bbu)$ the nodes selected by LGS for a vector $\bbu$ of per-link utilities, then our solution to Problem~\ref{P:main} is of the form
\begin{equation}\label{E:form}
\hat{\boldsymbol{v}}_{\mathrm{GCN}} = c(\mathcal{G}, \mathbf{q}, \mathbf{r}) = \mathrm{LGS}(\ccalG, \Psi_{\ccalG}(\bbS; \bbomega)).
\end{equation}
In \eqref{E:form}, $\Psi_{\ccalG}$ is an $L$-layered GCN defined on the conflict graph $\ccalG$, $\bbS$ is a matrix collecting the features for all $v \in \ccalV$, e.g., $\bbS=\left[\bbq, \bbr\right]$, and $\bbomega$ is the collection of trainable parameters of the GCN.
The downstream architecture of our solution is illustrated in Fig.~\ref{fig:system}.
At the system level, the GCN observes the state and topology of the network and generates per-link utilities as the expected returns of scheduling each link, based on which the LGS selects an independent set that \update{seeks to maximize} the total return as a schedule. 

Formally, by defining the output of an intermediate $l$th layer of the GCN as $\bbX^l \in\reals^{|\ccalV|\times g_{l}}$ with $\bbX^0 = \bbS$ and $\bbu_{\mathrm{GCN}} = \bbX^L $, 
we have that the expression for the $l$th layer of the GCN:
\begin{equation}\label{E:gcn}
	\mathbf{X}^{l} = \sigma\left(\mathbf{X}^{l-1}{\bbTheta}_{0}^{l}+\bbcalL \mathbf{X}^{l-1}{\bbTheta}_{1}^{l}\right), l\in\{1,\dots,L\},
\end{equation}
where $\bbcalL$ is the normalized Laplacian of $\ccalG$, ${\bbTheta}_{0}^{l}, {\bbTheta}_{1}^{l} \in \mathbb{R}^{g_{l-1} \times g_{l}}$ are trainable parameters, and $\sigma(\cdot)$ is the activation function. 
The activation functions of the input and hidden layers are selected as leaky ReLUs whereas a linear activation is used for the output layer. 
The output dimension is configured as $g_{L}=1$, so that $\bbu_{\mathrm{GCN}}$ is a vector.
The trainable parameters $\bbomega$ in~\eqref{E:form} correspond to the collection of ${\bbTheta}_{0}^{l}$ and ${\bbTheta}_{1}^{l}$ for all $L$ layers.

The output $\bbu_{\mathrm{GCN}}$ of the GCN is used as the per-link utilities in LGS. 
In general, the LGS algorithm builds an estimate $\hat{\boldsymbol{v}}_{\mathrm{Gr}}$ by iteratively adding vertices with the largest utility in their neighborhoods to the solution set, and then excluding them and their neighbors from the residual graph:
\begin{subequations}\label{E:lgs}
\begin{align}
	\hat{\boldsymbol{v}}_{\mathrm{Gr}} \gets \hat{\boldsymbol{v}}_{\mathrm{Gr}}\cup \Big\{v \Big| u(v) > \max_{v_i \in \mathcal{N}(v)} u(v_i) ,\; \forall\; v \in \ccalV' \Big\}\;, \label{E:lgs:set}\\
	\ccalG'(\ccalV',\ccalE') \gets\ccalG'(\ccalV',\ccalE') \setminus \left(\hat{\boldsymbol{v}}_{\mathrm{Gr}}\cup\mathcal{N}(\hat{\boldsymbol{v}}_{\mathrm{Gr}})\right)\;,\label{E:lgs:residual}
\end{align}
\end{subequations}
where $\mathcal{N}(\cdot)$ represents all vertices that are neighbors to a vertex or vertex set, and initially $\hat{\boldsymbol{v}}_{\mathrm{Gr}}=\emptyset$ and $\ccalG'=\ccalG$.
The algorithm terminates when $\ccalG'$ is empty, and outputs $\hat{\boldsymbol{v}}_{\mathrm{Gr}}$ as the solution, which is guaranteed to be an independent set by~\eqref{E:lgs:residual}. 
In the case of a tie in \eqref{E:lgs:set}, the link with a larger initially-assigned identification number wins without additional information exchanges. 

\begin{figure}[t]
	\centering
	\vspace{-0.1in}
	\includegraphics[width=\linewidth]{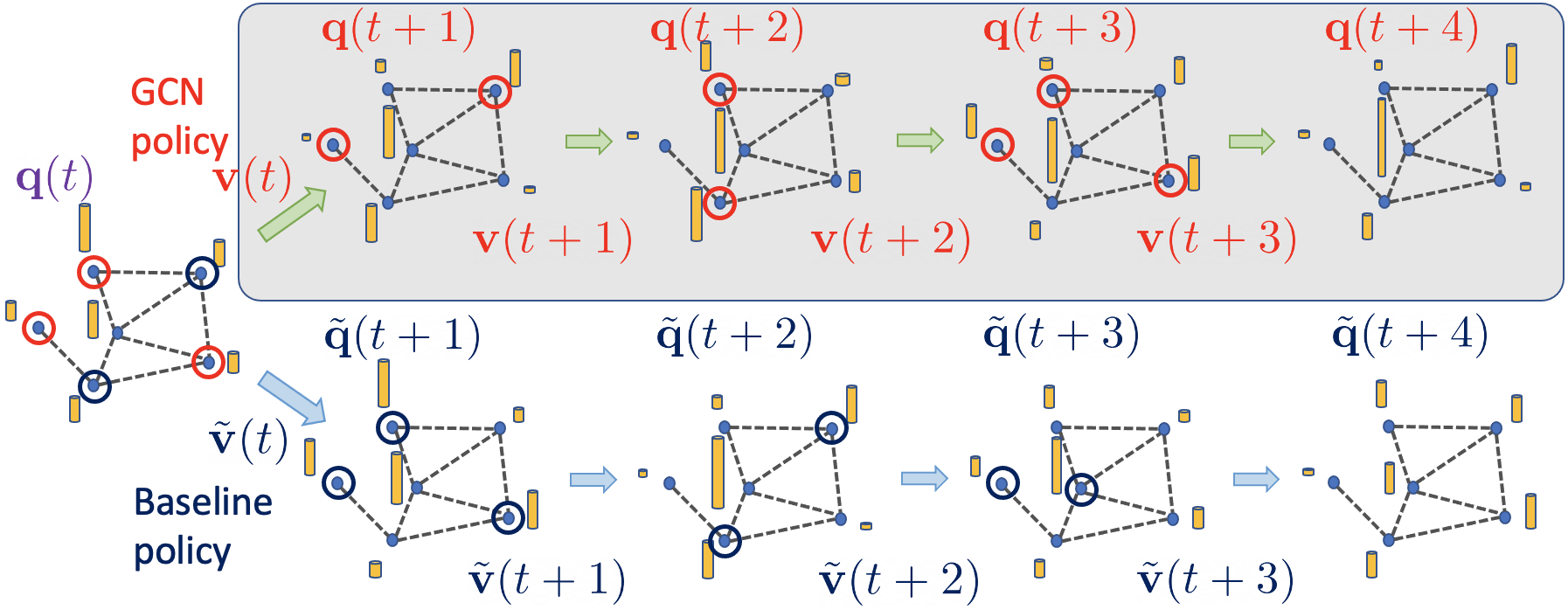}\vspace{-0.05in}
	\caption{In a 4-step lookahead scheduling, the GCN and baseline policies are executed for 4 steps from the network state at $t$, under identical environmental (arrivals and link rates) conditions. Whether the queues attained by the GCN improve upon the baseline is used to inform the quality of the scheduling decision at time~$t$. 
	}
	\label{fig:lookahead}
    \vspace{-0.1in}
\end{figure}

The local communication complexity (defined as the rounds of local exchanges between a node and its neighborhood) of the proposed scheduler is $\ccalO(L+\log|\ccalV|)$, where $\ccalO(\log|\ccalV|)$ is the average local complexity of LGS \cite{joo2012local}.
With~\eqref{E:gcn} and~\eqref{E:lgs}, $\hat{\boldsymbol{v}}_{\mathrm{GCN}}$ can be computed in a distributed manner, where the local computational and communication costs can be controlled by modifying the number of layers $L$ in the GCN.
Importantly, the logarithmic local communication complexity is a key aspect to promote scalability. 

It should be noted that the GCN in~\eqref{E:gcn} can generalize to networks of various sizes and topologies, and be deployed in a distributed manner. 
\update{Since $\ccalL$ is a local operator on $\ccalG$, $u_{\mathrm{GCN}}(v)$ can be computed locally at $v$ by neighborhood aggregation with $L$ rounds of local exchanges between $v$ and its neighbors \cite{kipf2016semi}.}
Furthermore, the application of LGS ensures the scheduled links $\hat{\boldsymbol{v}}_{\mathrm{GCN}}$ in~\eqref{E:form} form a valid independent set for \emph{any choice of trainable parameters} $\bbomega$.
However, we can train $\bbomega$ to minimize the objective in Problem~\ref{P:main}.

To train the parameters $\bbomega$ in the GCN, we first create \update{an interactive} virtual network environment, 
where the conflict graph $\ccalG$, packet arrivals $\bba$, and link rates $\bbr$, are drawn from (stationary) distributions of interest.
Next, experience tuples $\left(\ccalG(t), \bbS(t), \hat{\boldsymbol{v}}_{GCN}(t), \bbrho(t)\right)$, for $t\in\{0,\dots,T\}$ are collected from the GCN-based scheduler interacting with the virtual environment. The return vector $\bbrho(t)$ captures the relative performance of the GCN under training with respect to a baseline policy in a $K$-step lookahead scheduling from network state of $t$, as illustrated in Fig.~\ref{fig:lookahead}.
As a baseline, we use LGS based on the per-link utility given by $\bbu(t)=\bbq(t)\odot\bbr(t)$ \cite{joo2012local}, where $\odot$ denotes an element-wise product.
Formally, $\bbrho(t)$ is given by
\begin{equation}\label{E:reward}
	\bbrho(t) = \varphi\left(\frac{\sum_{k=1}^{K}\lVert\tilde{\bbq}(t+k)\rVert_1}{\sum_{k=1}^{K}\lVert\bbq(t+k)\rVert_1}\right) {\bbv}(t) + \bbu_{\mathrm{GCN}}(t)\odot\left[\mathbf{1}-\bbv(t)\right]\;,
\end{equation}
where ${\bbv}(t)$ is the indicator vector of schedule $\hat{\boldsymbol{v}}_{\mathrm{GCN}}(t)$, $\tilde{\bbq}(t+k)$ is the vector of queue lengths under the baseline policy $k$ steps after $t$, and $\varphi$ is an activation function that can be either linear $\varphi(x)=x$ or a Heaviside step function $\varphi(x)=H(x-1)$. 
Intuitively, whenever the GCN policy beats the baseline in a $K$-step lookahead scheduling (higher values of $\bbrho(t)$), the schedule $ \hat{\boldsymbol{v}}_{\mathrm{GCN}}(t)$ is encouraged for the network state at $t$.
Consequently, a root-mean-square loss is adopted to train our GCN. 
For a given experience tuple this is given by
\begin{equation}\label{E:gcn_loss}
\ell(\bbomega; \ccalG(t), \bbS(t)) = |\ccalV|^{-\frac{1}{2}} \lVert \bbu_{\mathrm{GCN}}(t)-\bbrho(t) \rVert_{2}.
\end{equation}
With the loss in \eqref{E:gcn_loss} and the collected experience tuples, we update the parameters $\bbomega$ of the GCN through batch training, employing the Adam optimizer and exponentially decaying learning rates. 

\section{Numerical experiments}
\label{sec:results}


\begin{figure}[t]
\centering
\vspace{-0.3in}
    \includegraphics[width=0.9\linewidth]{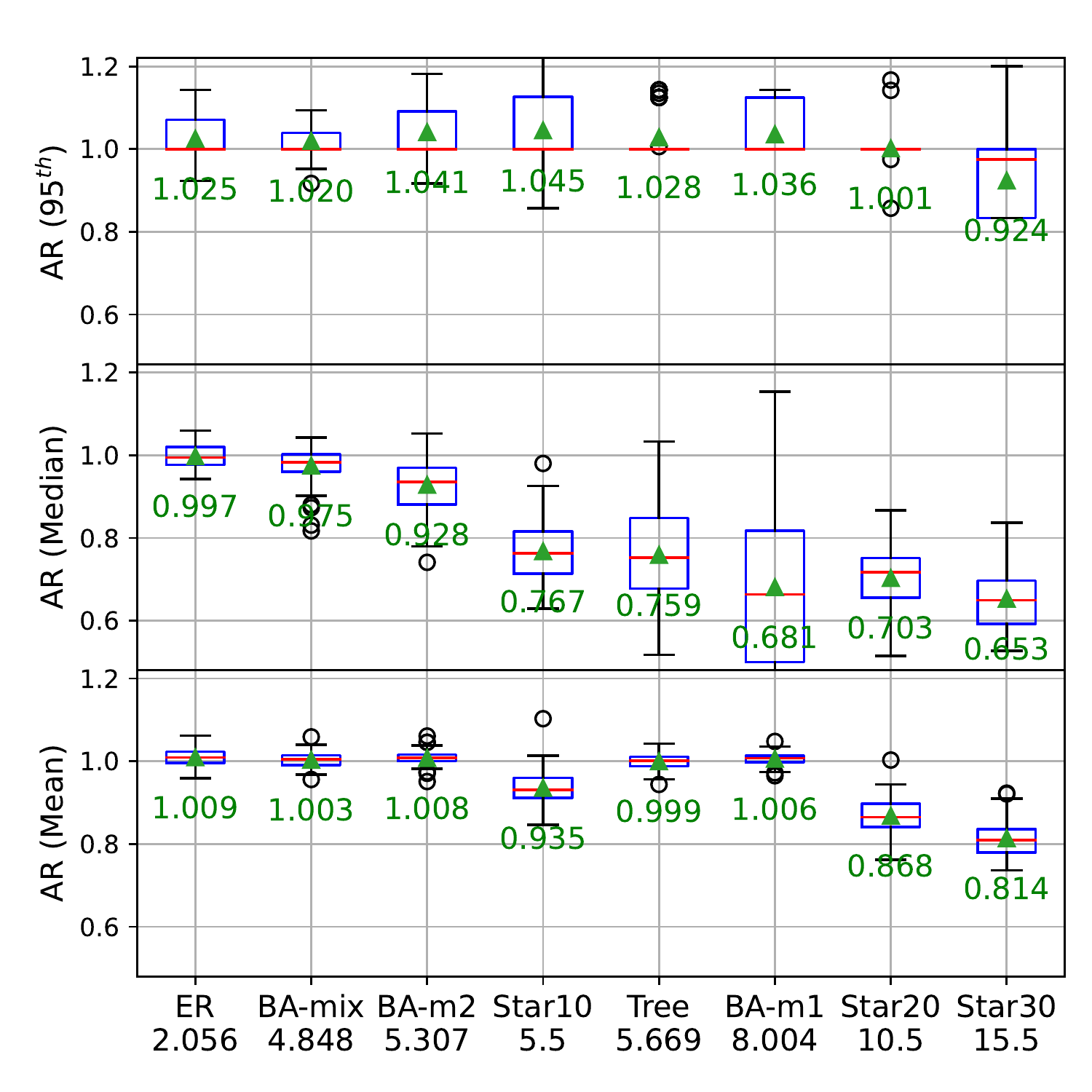}
    \vspace{-0.15in}
    \caption{The approximation ratios of GCN($L=1$)-based distributed scheduler relative to local greedy scheduler \cite{joo2012local} under various conflict graph configurations: (top) $95^{th}$ percentile, 
    (middle) median, and  
    (bottom) mean backlogs. 
    Smaller values refer to better delay.
    }   \label{fig:results:mean}
 \label{fig:results}    
 \vspace{-0.2in}
\end{figure}

We evaluate our GCN-based distributed scheduler in simulated wireless networks.
The simulated conflict graphs include star graph, and random graphs generated following the Erdős–Rényi (ER)~\cite{erdds1959random}, Barabási–Albert (BA)~\cite{Albert02}, and power-law tree models. 
\update{Intuitively, the ER conflict graphs seek to represent networks of uniformly distributed users with identical transmit power (i.e., unit-disk interference zones).
The star and BA conflict graphs arise from networks with several connected components and links of heterogeneous transmit power, e.g., macrocells surrounded by microcells and D2D links.
The power-law trees represent wireless backhaul networks where some links have interfering neighbors.}

We adopt the following configurations for conflict graphs:
\textit{Star$X$} (star graph with $V=X+1$), \textit{BA-m$X$} (BA graph with $V=70,m=X$), \textit{BA-mix} (BA graphs with $V\in\{100,150,\dots,300\}$ and $m\in\{2, 5, 10, 15, 20\}$), \textit{ER} (ER graphs with $V=50,p=0.1$), and \textit{Tree} (power-law tree with $V=50,\gamma=3$),
where $V=|\ccalV|$ is the number of nodes, $m$ is the number of edges that each new node forms during the preferential attachment process for the BA model, $p$ is the probability of edge-appearance for the ER model, and $\gamma$ is the exponent for the power-law tree model.
Link rates $r_v(t)$, defined as the number of packets that can be transmitted through link $v$ at time $t$, are independently drawn from a normal distribution $\mathbb{N}(50, 25)$, and clipped to $\left[0,100\right]$, 
to capture a lognormal fading channel~\cite{Mousavi17lte}.
The exogenous packets at each source user follow a Poisson arrival with a prescribed arrival rate $\lambda$.
We define network traffic load as $\mu = \lambda/\mathbb{E}(\bbr)$.
For each configuration of a graph model, we generate $100$ scheduling instances, each contains realizations of a conflict graph and the random processes of arrivals and link rates for $T=64$ time slots.
A 1-hop flow is generated for each link in the network. 
In a scheduling instance, each tested scheduler is tested under identical realizations of random processes.

\begin{figure}[t]
	\centering
	\vspace{-0.15in}
	\includegraphics[width=0.85\linewidth]{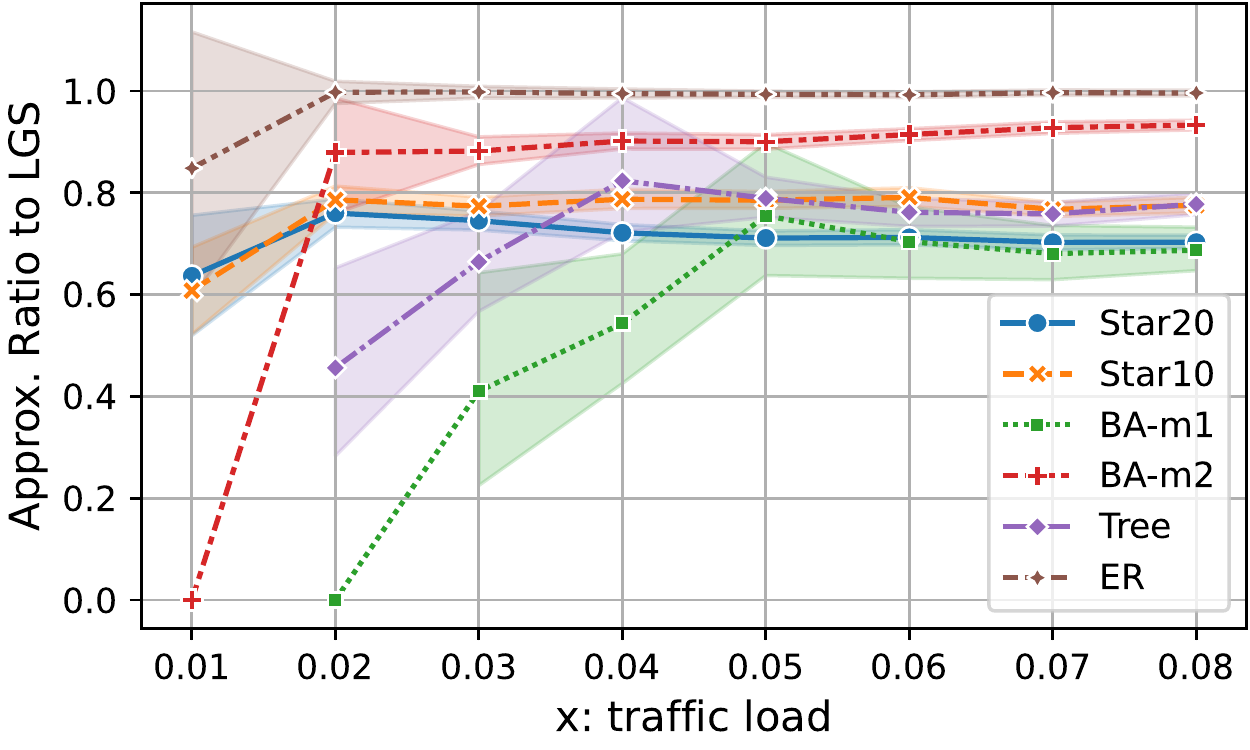}\vspace{-0.1in}
	\caption{The approximation ratios of the median backlogs of GCN($L=1$)-based scheduler w.r.t. the baseline by traffic load.}
	\vspace{-0.1in}
	\label{fig:medbyload}
	\vspace{-0.1in}
\end{figure}

A single-layer GCN ($L=1$) is evaluated. 
The baseline for training and evaluation is the vanilla LGS with a utility function of $\bbu(t)=\bbq(t)\odot\bbr(t)$~\cite{joo2012local}.
The feature matrix is set to be the same as the baseline utility $\bbS(t)=\left[\bbu(t)\right]$. 
Through trial-and-error, we train the GCN on scheduling instances with mixed conflict graphs of $80\%$ \textit{Star30} and $20\%$ \textit{BA-m2} for effectiveness and generalizability. 
The GCN-DQL settings include $\varphi(x)=H(x-1)$, $K=5$, a batch size of 64 for experience replay, and 6000 episodes.\footnote{Training typically takes 3-4 hours on a workstation with a specification of 16GB memory, 8 cores, and Geforce GTX 1070 GPU. The source code is published at \url{https://github.com/zhongyuanzhao/gcn-dql} }

We collect the mean, median, and tail ($95^{th}$ percentile) queue lengths (backlogs) over time and links in the wireless networks, under the GCN-based distributed scheduler and the baseline, in light-to-moderately loaded traffics, $\mu=0.01,\dots,0.08$. 
The boxplots of approximation ratios (AR) for the $95^{th}$ percentile, median, and average backlogs by graph model, of the GCN-based scheduler to the baseline (smaller than $1$ denotes improvement) with $\mu=0.07$ are illustrated in Figs.~\ref{fig:results:mean} (top to bottom), respectively, where the means are in green text and marked by green triangles.
The peak to average degree ratio (written under each graph annotation in Fig.~\ref{fig:results:mean}) measures the graph centralization. 
Compared to the baseline, our GCN-based scheduler can reduce the average and median backlogs, especially when the confict graph is more centralized and the central links are more likely to be congested by the memoryless baseline. 
On star graphs, the average and median backlogs are respectively reduced by $6.5\sim18.6\%$ and $23.3\sim34.7\%$.
On more complex topologies, the mean backlogs are the same as the baseline, while the median backlogs are reduced proportionally to the graph centralization, e.g. from $2.5\%$ on BA-mix to $32.9\%$ on BA-m1.
The tail backlogs of GCN-based scheduler are increased by $0.1\sim 4.5\%$ on average, while the median values are the same as the baseline, except on Star30 where tail backlogs are improved in both mean and median values. 
With a different baseline utility of $\bbu(t)=\min(\bbq(t),\bbr(t))$, the average backlogs on BA and Tree graphs can also be improved by the GCN. 
\update{Considering sojourn time~\cite{hai2018delay} as the baseline utility, the average sojourn time of a packet can be improved by $1.6\sim4.6\%$ on star graphs.} 
These results show that our approach can improve the delay of the majority of the links over the baseline, at the cost of slight increase of the tail delay.
The improvement of the median backlogs brought by the GCN-based scheduler is consistent under different traffic loads, as illustrated in Fig.~\ref{fig:medbyload}, which shows its good generalizability to network traffic conditions.

\vspace{-2mm}
\section{Conclusions}
\label{sec:conclusions}
\vspace{-2mm}

We presented a GCN-based distributed and scalable scheduler to improve the delay performance in wireless networks by combining the efficiency of a local greedy scheduler and the ability of GCNs to encode the network state and topological information. 
The proposed scheduler markedly improves upon classical baselines, especially in wireless networks with \update{several connected components and heterogeneous transmit power}, and shows good generalizability over graph types, graph sizes, and traffic loads.

\vfill\pagebreak



\bibliographystyle{ieeetr}
\bibliography{strings,refs}

\end{document}